\documentclass{article}
\usepackage{spconf,amsmath,graphicx}
\usepackage[icelandic,english]{babel}
\usepackage[T1]{fontenc}
\usepackage[dvipsnames]{xcolor}
\usepackage[normalem]{ulem}

\title{Manual Post-editing of Automatically Transcribed Speeches from the Icelandic Parliament - Althingi}
\name{Judy Y. Fong, Michal Borsky, Inga R. Helgad{\'o}ttir, Jon Gudnason \thanks{This project was made possible through the support of Althingi's information and publications departments. The authors would like to thank Solveig K. J{\'o}nsd{\'o}ttir and  \TH orbj{\"o}rg {\'A}rnad{\'o}ttir for their valuable help, Ingvi St{\'i}gsson for handling the technical aspects of the test at Althingi, as well as the editors, for reading over the ASR transcriptions, correcting them, timing the process and giving valuable comments.}}

\address{Reykjavik University \\
		Language and Voice Lab \\ 
        Menntavegur 1, 101 Reykjavik, Iceland}

\begin{document}
\maketitle
\begin{abstract}
The design objectives for an automatic transcription system are to produce text readable by humans and to minimize the impact on manual post-editing. This study reports on a recognition system used for transcribing speeches in the Icelandic parliament - Althingi. It evaluates the system performance and its effect on manual post-editing. The results are compared against the original manual transcription process. 239 total speeches, consisting of 11 hours and 33 minutes, were processed, both manually and automatically, and the editing process was analysed. The dependence of word edit distance on edit time and the editing real-time factor has been estimated and compared to user evaluations of the transcription system. The main findings show that the word edit distance is positively correlated with edit time and a system achieving a 12.6\% edit distance would match the performance of manual transcribers. Producing perfect transcriptions would result in a real-time factor of 2.56. The study also shows that 99\% of low error rate speeches received a medium or good grade in subjective evaluations. On the contrary, 21\% of high error rate speeches received a bad grade. 
\end{abstract}


\section{Introduction}

In the last 5 years, automatic speech recognition (ASR) technology has advanced enough to be used in real-life applications. Recognition technology has been used extensively to transcribe speeches for large languages such as English, German or Spanish~\cite{miro2015efficiency,munteanu2008collaborative,kolkhorst2012evaluation}. These systems are often composed of an ASR module to produce audio-to-text transcription and several natural language processing modules to improve text formatting. The main issue, however, is that neither module performs with perfect accuracy so manual post-processing is needed to produce final transcriptions.

The system for automatically transcribing university lectures in Spanish~\cite{miro2015efficiency} compared three post-processing approaches: one involving automatic corrections, another using lecturer corrections, and a third using a mixture of both. The system was tested on twenty lectures and the \mbox{\it WER} was compared to a real-time factor of the post-editing time versus the total duration of the lecture. The authors conclude that the edit time is directly correlated with the transcription accuracy, but the relationship between the real-time factor and \mbox{\it WER} was weak, perhaps due to the low \mbox{\it WER} range produced by the ASR. In the English transcription system~\cite{munteanu2008collaborative}, the challenge of achieving a low error rate in transcribing university lectures was handled using collaborative editing. The authors' findings conclude that correcting transcripts with $\mbox{\it WER}$ lower than 25\% increases the editing effort. The transcription errors for lectures in German~\cite{munteanu2008collaborative} were corrected using student edits and this error correction was studied. During the transcription process they noted that their ASR made errors caused by uncommon and non-German terms in the lectures. Their analysis showed that the corrections of inexperienced editors tend to bring a high WER down to about 25\%, corroborating the findings of~\cite{munteanu2008collaborative}. 

Evaluation of post-editing transcribed speech was studied in~\cite{sperber2017transcribing}, where authors observe a strong variation in editing accuracy and speed among editors. Authors also note that low $\mbox{\it WER}$ transcripts require advanced editing strategies to achieve error rate improvements comparable to improvements for high $\mbox{\it WER}$ transcripts. Different transcription strategies were compared in~\cite{Sperber2016}; namely a fully manual post-editing of ASR transcripts and confidence-enhanced post-editing of ASR transcripts. The authors conclude that post-editing automatic transcripts results in more accurate and faster transcripts, when compared to manually transcribing from scratch. This conclusion was further corroborated in~\cite{Miro2018}, which dealt with automatic subtitling of videos.

This paper evaluates an ASR system in the context of transcribing speeches for the Icelandic parliament - Althingi. The system has only recently been developed for Icelandic~\cite{gudhnason2012almannaromur,Helgadottir2017corpus,gudhnason2017building,NIKULASDOTTIR2018}, and is now being incorporated into the transcription process of Althingi. The current manual procedure is done in two stages: an initial manuscript is obtained from a contracted transcription service, which is then post-edited by in-house specialists. The main objective of the current project is to replace the initial manual transcription process with an automatic speech recognizer. This is the first time an ASR system is used as a core component in transcription for the Icelandic language, and the purpose of this paper is threefold; to introduce the evaluation procedure, to present the first measurements of the manual post-editing and to report on the performance of the system.

\section{Transcription System for Althingi}

\begin{figure}[h]
  \centering
  \includegraphics[width=\linewidth]{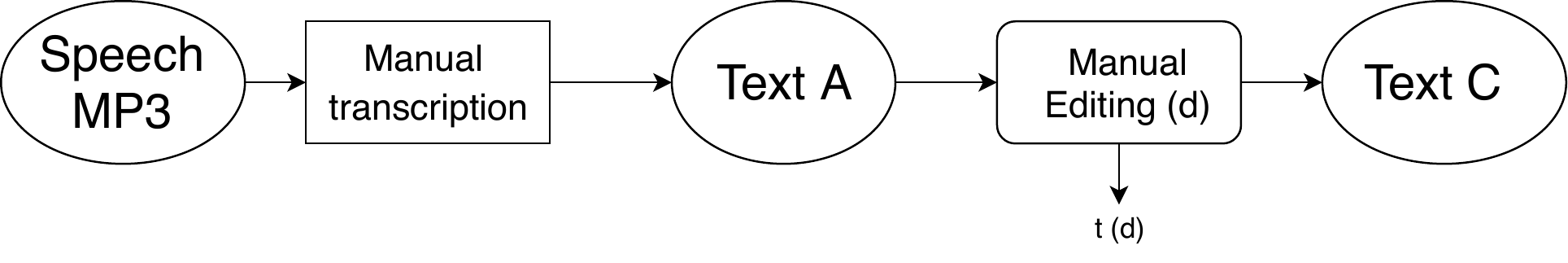}
  \caption{Diagram of the transcription and editing process for Althingi. t(d) indicates editing time.}
  \label{fig:althingi_workflow}
\end{figure}

The current transcription procedure for the Icelandic parliament is done in two stages, illustrated in Figure \ref{fig:althingi_workflow}. The speeches are first created in the Althingi document management system, Documentum, as XML documents, with only the speech meta-data and a link to the speech in the MP3 format. Then, in the manual transcription stage the transcribers listen to the audio and transcribe the speech into the XML document (Text A). The initial transcript is meant to reflect the spoken record as accurately as possible but the transcribers might also enrich the text with minor changes. For example, they might add in different formatting for poems and remove repetitions. Next, in the manual editing stage the XML speech document is sent to the editors who modify the speech to be fit for publication and record their editing time. It is common that an editor corrects transcription errors, fixes grammar or enriches the text with context to make the parliamentary record clearer. Finally, the speeches (Text C) are published to their website.


The main objective of the current project is to replace the first stage, manual transcription, with an automatic speech recognizer. Before the experiment, the in-house specialists gave suggestions regarding relevant data to gather and discussed the important differences between the ASR and manual transcriptions. For the experiment, the manual transcription and ASR transcriptions were done in parallel. With the intent of using Text A as reference material, only the ASR transcriptions received manual post-editing. The experiment was performed for a week; on the first day, only the first stage was tested, to ensure the integration worked as intended. For that week the Icelandic parliament was in session for four days. It is from the last three days that this data was gathered.

\subsection{The ASR system}
The details about the preparation of the ASR training data and the development of the ASR can be found in~\cite{Helgadottir2017corpus}. The acoustic model is a deep neural network, based on a recipe developed for the Switchboard corpus\footnote{https://github.com/kaldi-asr/kaldi/blob/master/egs/swbd/s5c/\\local/chain/tuning/run\_tdnn\_lstm\_1e.sh}, using the Kaldi ASR toolkit~\cite{povey2011kaldi}. It is a sequence trained neural network based on lattice-free MMI~\cite{povey2016purely}. It consists of seven time delay deep neural network layers~\cite{waibel1989phoneme} and three long-short term memory layers~\cite{sak2014long}. The network takes 40 dimensional LDA feature vectors and a 100 dimensional i-vector as input.
Two n-gram language models were trained using the KenLM toolkit \cite{heafield2011kenlm}. The first one is a pruned trigram model, used in the decoding. The other one is a 5-gram language model, trained on the total parliamentary text set, 55M tokens, and is used for re-scoring decoding results.
The lexicon is based on the pronunciation dictionary from the Hjal project \cite{rognvaldsson2003icelandic}, available at M{\'a}lf{\"o}ng\footnote{http://www.malfong.is}. We added words from the language model training data, which appeared three or more times, with some constraints, resulting in roughly a dictionary containing 200k words. Inconsistencies in the pronunciation dictionary were also fixed.
The WER of the ASR, before any post-processing is done, is $9.63\%$ on the test set, using 1500 hours of parliamentary speeches and corresponding text, for training. In real life, this number is going to be higher, partly because of imperfect punctuation reconstruction and disparate casing of many words in our texts, and partly because the ASR test set had been manually cleaned to better match the audio.

\subsection{Automatic post-processing}
The ASR returns a stream of words with no punctuation or formatting. Since the purpose of the system is to publish parliamentary speeches, human readability needs to be factored into the final transcription. The OpenGrm Thrax Grammar Development tool~\cite{tai2011thrax,roark2012opengrm} was used to compile grammars into weighted finite-state transducers, in order to denormalize numbers and abbreviations, according to parliamentary conventions. 

The Punctuator toolkit~\cite{tilk2016bidirectional} is used to restore punctuations in the text, specifically periods, question marks and colons. There are no clear rules for the use of commas in Icelandic, making learning their position difficult. Therefore, no commas are added to the ASR transcripts. Punctuator is a bidirectional recurrent neural network model with an attention mechanism. It can both be trained on punctuation annotated text only, or additionally, take in pause annotated text. Both versions were tested, with the text-only training giving better results, an overall $F1$-score of $86.7$ versus a score of $83.7$ for the two stage training. These errors are obtained on well structured text and are likely higher in the automatically transcribed speeches. The training text set contains roughly 50M words. The development and test set contain 114k and 111k words, respectively. The pause annotated text set contains 1.3M words. The pause annotated development and test sets are each around 81k words. The pause information is obtained from existing alignment lattices, from earlier data preparations before the ASR training.

Apart from punctuation, formatting also plays a large factor in human readability. Therefore, Thrax grammar rules are used to capitalize the start of sentences and to collapse expanded acronyms. They are also used for other small formatting, such as timestamps, regulations, time intervals, and websites. Another important task for long texts is adding paragraph insertions. Currently, a new paragraph is only started whenever the speaker of the house is addressed.

\subsection{Integration with the Althingi system}

The ASR needs to connect with the Althingi servers in four different instances. This is enabled for the first three occurrences via a representational state transfer application programming interface (RESTful API). The API first takes in the timestamp of when the speech ends through a GET request. Then, using the ending timestamp, the ASR server queries Althingi's metadata server to obtain the timestamp of when the speech started. With the two timestamps, the ASR server queries Althingi's audio server for the audio segment, which the ASR server then downloads. The rest of the experiment is semi-automatic. With the ASR, the audio is then transcribed. Next, the ASR TXT document (Text B of Figure \ref{fig:user_test_1b}) is batched and wrapped in the speech metadata as well as XML tags. Finally, they are copied from the ASR server and manually entered into the Documentum editor queue. After the speeches are post-edited (Text D), they are posted onto the Althingi website\footnote{http://www.althingi.is}.

Currently, the ASR is housed on its own server and interacts with the rest of the Althingi servers through the RESTful API. The ASR is built on a Ubuntu 16.04 server with 4 CPUs and 16 GB of RAM. The number of parallel transcriptions are limited by the number of CPUs within the server. During the test, 3-4 speeches were processed in parallel because members of parliament tended to deliver speeches faster than the ASR could transcribe.

\subsection{ASR integration concerns}

Since this integration was only for the experiment, not permanent, the ASR speeches needed to be delivered to the editors while still keeping the existing transcription procedure intact. In order to manage this, automatically transcribed speeches were manually entered  into Documentum. To accomplish the goal of keeping the test procedure separate but integrated, Althingi put the ASR speeches in their own separate folder which was then integrated with the normal procedure at the post-editing stage through the Documentum lifecycles. The in-house specialists' queue only showed the ASR transcriptions.

Despite familiarity with the technical details, a deeper understanding of the Althingi speech publishing procedure was needed. Thus, several of their in-house specialists were requested to give valuable insight on important details in the post-editing procedure which could not be gleaned from data. For example, the idea of inserting a new paragraph when parliamentarians address the speaker of the house as it will usually signal a change in topic originated from these specialists.

\section{Methodology}
 
\begin{figure}[h]
  \centering
  \includegraphics[width=\linewidth]{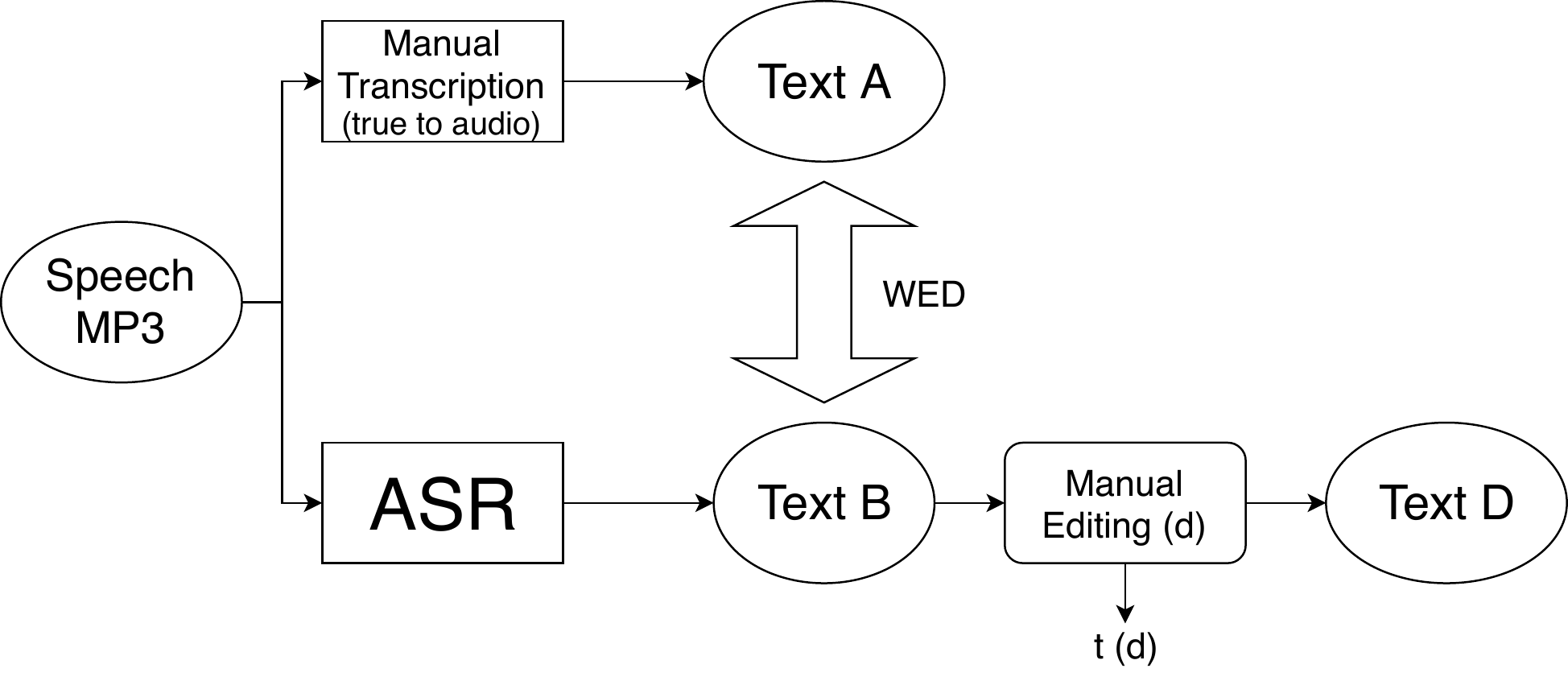}
  \caption{Diagram of the transcription and editing process for the experiment. WED is the word edit distance of Text B from Text A. t(d) is the edit time [s] the editors take to post-process the speech. }
  \label{fig:user_test_1b}
\end{figure}

The primary objective for this experiment is to discover the impact of switching from manual transcriptions to automatic transcriptions on the Althingi publication department's work. Figure \ref{fig:user_test_1b} illustrates the experimental setup. First, the speech segment is sent to both the manual transcription stage and the ASR stage. Wherein, Text A and Text B are created. Then, only Text B is sent to the editor queue for the in-house specialists to post-edit and produce the final transcription, Text D.

Over a three day period, the Icelandic parliament delivered 279 speeches. However, at the conclusion of the experiment, 35 speeches still hadn't been processed by the publications department, and 5 speeches were duplicates. Therefore, only 239 speeches could be analysed.  The data collected includes the following: speech length, word count, edit time, editor feedback and calculations of the subsequent measures. The system was evaluated using the following measures: 1) word edit distance (\mbox{WED}) [\%], 2) edit time per word (\mbox{ET/W}) [s/w], and 3) real-time factor (\mbox{RTF}) [-]. All three measures reflect on an editor's effort in processing a transcribed speech. The \mbox{WED} was calculated using the following formula:

\begin{equation}
WED = \frac{S+I+D}{N}*100
\end{equation}
\noindent
where \mbox{\it S}, \mbox{\it I}, \mbox{\it D}, \mbox{\it N} is the number of substitutions, insertions, deletions and total words respectively, obtained by aligning the texts. This formula is identical to \mbox{\it WER}, but since it also reports on the edit distance between transcription and final text, the term word edit distance is preferred. The \mbox{RTF} was computed as the edit time in seconds divided by the speech length in seconds.

For Text A, the transcribers were asked to transcribe the audio as true as possible, leaving in speaker errors, in order to get good reference texts. Since this is contrary to the work they normally do, some of the texts were not true-to-audio. The manual transcriptions tended to have small corrections since repetitions are removed, badly structured sentences are corrected, and three words common to parliamentary speeches are abbreviated. In addition, they also contained spelling mistakes and word substitutions due to malformed speech. Despite these flaws, Text A is still the better reference than Text D when estimating errors as both automatic and manual transcription aim to produce audio-to-text transcription. However, it is true that the DB alignment better reflects the work the editors do to make an automatically transcribed text publishable. Hence, one would expect the \mbox{\it WED} between Text B and Text D to better explain ET/W than the edit distance between Text A and Text B. The DB results are obtained under the verification and guidance of the AB results. 

Editors gave feedback in the form of comments and grades for the whole system and on individual speeches. While recording the edit time for a speech the editors were also asked to grade and comment on the speech based on their own perceptions. There were no guidelines other than giving the speeches a grade (Good, Medium, or Bad). Not giving them guidelines better simulates their day-to-day feelings. After the experiment they filled out a short survey with their evaluations of the current procedure versus the inclusion of the automatic transcription system.

In the succeeding week speeches were edited with the procedure illustrated in Figure~\ref{fig:althingi_workflow} to produce the results for the fully manual procedure,  referenced as 'Fully Manual' later in the the text. They lend insight into the speed of the fully manual transcription process.

\section{Results}

The ultimate goal of the automatic transcription system is to replace human transcribers, so the obvious benchmark to compare against are the results for the fully manual transcription process. This would include matching the \mbox{\it ET/W} and \mbox{\it RTF}, but not necessarily the word edit distance.  

\begin{table}[!h]
  \caption{Influence of the fully manual and automatic transcription on editing effort.}
  \label{tab:summary}
  \centering
  \begin{tabular}{ l l l }
    \hline
	  & \textbf{ET/W} [s/w] & \textbf{RTF} [-]    \\
    \hline
   Fully Manual & 1.32 $\pm$ 0.51 & 2.66 $\pm$ 1.05 \\
   Automatic & 1.52 $\pm$ 0.53 & 3.26 $\pm$ 1.24 \\
  \end{tabular}
\end{table}

The DB alignment results are summarized in Table~\ref{tab:summary}. The analysis shows that the automatic process under-performs when compared to the fully manual process. The \mbox{\it ET/W} is higher by 0.20 s/w, and the \mbox{RTF} by 0.60. The initial hypothesis was that the edit distance would be the main factor affecting the edit time and that the higher the distance, the higher the edit time. Also, that both RTF and ET/W would positively correlate with \mbox{\it WED}. In order to confirm this hypothesis, the linear correlation analysis was performed to model the dependence of \mbox{\it ET/W} and \mbox{\it RTF} on \mbox{\it WED}. Also, the Pearson's correlation coefficient (PCC) between the two variables was computed. The results are as follows:

\begin{itemize}
\setlength\itemsep{0.05em}
\item ET/W = {\it 0.019} * WED + {\it 1.08}
\item RTF = {\it 0.03} * WED + {\it 2.56}
\item PCC(WED,ET/W) = 0.33
\item PCC(WED,RTF) = 0.22
\end{itemize}

These are the observations from this analysis: \textbf{1)} reducing the ET/W to the manual level would require lowering the \mbox{WED} to 12.6\%, \textbf{2)} producing perfect automatic transcriptions can only outperform manual transcription RTF by $\approx4\%$, since, the cost of reading through the transcription, independent of any errors, far outweighs the impact of errors, \textbf{3)} the correlation between the variables is low, indicating that WED might not be the best predictor of editing efforts.

The following assessment focused on system performance in terms of several error types: ASR, punctuation, capitalization and abbreviation mismatches. The analysis showed that the majority of transcription errors occur due to the ASR and wrong punctuation by far, followed by capitalization, and abbreviation respectively. As a consequence, an improvement to the ASR appears to be of the highest priority. However, in the post-experiment survey the editors frequently commented on inaccurate punctuation which prompted a singling out of punctuation from other errors and to study its affect on editing time. This approach also helped answer the question of whether a certain type of error takes more time to correct than others.

The following results distinguish between \mbox{WED} with punctuation (WED\_wp) and without punctuation (WED\_wop). The data was categorized into two groups, highs (H) and lows (L), representing speeches with high or low WED, with and without punctuation. Table~\ref{tab:avgWED_DB} summarizes average edit distance values for DB alignment, and Table~\ref{tab:avgWED_AB} for AB alignment. High was chosen as $\mu+0.25*\sigma$ and low as $\mu-0.25*\sigma$. The assumption is that speeches with both high WED\_wp and low WED\_wop will give insight into the effect of punctuation errors on editing time. Likewise, assuming speeches with low WED\_wp and high WED\_wop gives insight into the effect of all other errors on the editing time. The H-H group provides an opportunity to study deficiencies of our system that need to be addressed. The L-L group, on the other, gives an impression of the current performance ceiling. The table shows that excluding punctuation from transcripts improved the WED metric by $\approx6\%$ for both AB and DB alignment in absolute terms, likely due to the removal of the start of sentence capitalization errors introduced by the punctuation module. However, the editors always get transcripts with punctuation, so the values of WED\_wp are more relevant to editing effort. The same assumption is true with regards to DB alignment.

\begin{table}[!h]
  \caption{The average WED [\%] for DB alignment and the cut-off points for the High and Low groups.}
  \label{tab:avgWED_DB}
  \centering
  \begin{tabular}{ l l l }
    \hline
	  & \textbf{WED\_wp} & \textbf{WED\_wop}     \\
    \hline
   Average & 24.20 $\pm$ 9.56 & 19.58 $\pm$ 9.59  \\
   High & $>$ 26.59 & $>$ 21.97 \\
   Low  & $<$ 21.81 & $<$ 17.18 \\
  \end{tabular}
\end{table}

\begin{table}[!h]
  \caption{The average WED [\%] for AB alignment and the cut-off points for the High and Low groups.}
  \label{tab:avgWED_AB}
  \centering
  \begin{tabular}{ l l l }
    \hline
	  & \textbf{WED\_wp} & \textbf{WED\_wop}     \\
    \hline
   Average & 20.12 $\pm$ 5.80 & 14.35 $\pm$ 4.78  \\
   High &   $>$ 21.57 & $>$ 15.45 \\
   Low &  $<$ 18.67 & $<$ 13.15\\
  \end{tabular}
\end{table}

The average values of \mbox{ET/W}, \mbox{RTF} and speech count for DB alignment are summarized  in Table~\ref{tab:resultsDB}. Singling out the L-L group also showed that when the transcription system is doing as well as it can, the corresponding \mbox{ET/W} (1.36) is comparable to the effort for manual transcripts (1.32). This corresponds to a 3\% relative difference. The relative difference for RTF reached about 10\%. On the other hand, the relative differences between the H-H group and the Fully Manual process  reached 22.8\% and 31.9\% for ET/W and RTF respectively. The direct comparison between H-H and L-L shows a similar dramatic increase in both measures, clearly proving that the higher the edit distance, the higher the editing effort. The immediate concern for in-between groups is a lack of data to draw statistically significant conclusions. 

\begin{table}[th]
   \caption{ET/W and RTF results for selected groups for DB alignment.}
   \label{tab:resultsDB}
   \centering
   \begin{tabular}{ l l c c c }
     \hline
 	\textbf{WED\_wp} & \textbf{WED\_wop} & \textbf{\# speeches}& \textbf{ET/W} & \textbf{RTF} \\
     \hline
     High  & High & 84 & 1.71 & 3.51  \\
     High  & Low  & 1 & 1.14 &  1.6   \\
     Low   & High & 0 & - &  - \\ 
     Low   & Low  & 101 & 1.36 &  2.96  \\
   \end{tabular}  
\end{table}

The average values of \mbox{\it ET/W}, \mbox{\it RTF} and speech count for AB alignment is summarized in Table~\ref{tab:resultsAB}. The general trends for the H-H and L-L groups are identical to the DB alignment, confirming our initial hypothesis. This time, however, there were some points for in-between categories. The data shows that high WED\_wop leads to higher edit times. Therefore, from the numbers themselves one might conclude that punctuation errors take less time to correct than other errors, most being ASR-based errors. This observation is further supported by fixing WED\_wop as H and changing WED\_wp. But the H-L cluster shows lower \mbox{\it ET/W} and \mbox{\it RTF} than even the manual process, which indicates that the punctuation errors are less severe than the other errors. The main issue, however, with these conclusions are too few data point so further experiments are needed to confirm or deny the validity of this finding.  

\begin{table}[th]
   \caption{ET/W and RTF results for selected groups for AB alignment.}
   \label{tab:resultsAB}
   \centering
   \begin{tabular}{ l l c c c }
     \hline
 	\textbf{WED\_wp} & \textbf{WED\_wop} & \textbf{\# speeches}& \textbf{ET/W} & \textbf{RTF} \\
     \hline
     High  & High & 73 & 1.80 & 3.86      \\
     High  & Low  & 4 & 1.25  & 2.14      \\
     Low   & High & 5 & 1.43  & 3.02      \\ 
     Low   & Low  & 91 & 1.37 & 2.90      \\
   \end{tabular}  
\end{table}

Part of the experiment was also to obtain subjective evaluations of the system from the editors. The editors' independent evaluation of most speeches showed that for the 234 graded speeches 26 were graded as Bad, 105 as Medium, and 103 as Good. Comparing the grades to the H-H/L-L groups, shows that 68\%/70\% of the L-L speeches were graded as Good and only 2\%/1\% as Bad, for Texts AB and Texts DB, respectively. In the H-H group 27\%/21\% of the speeches were graded as Bad and 18\%/18\% as Good, for Texts AB and Texts DB, respectively. These numbers are highly subjective and vary between editors but give an otherwise hard to obtain insight on the how in-house editors' opinions line up with the other data.

Reading the comments the editors wrote about the speeches in these two groups show some differences. More of the speeches in the H-H groups have comments and the comments are longer. Most prominent are complaints about word substitutions and deletions, as well as incorrect punctuation. For these speeches the editors often mention that the speaker is hard to understand. Comments in the L-L group are fewer. However, complaints about incorrect capitalization are more prominent.

Multiple factors also contributed to a higher edit time and WED: dealing with Roman numerals, differences in repetitions or incorrect capitalization. But the edit time alone also had many factors contributing to it other than WED. Editors often formatted the text, such as splitting the speech into paragraphs. Sometimes, editors researched references to bills mentioned in the speeches. Other times, they had to pay close attention to the audio. Members of parliament would sometimes mention named-entities from different languages, which is outside of the scope of a monolingual ASR.





\section{Conclusion}

This paper evaluated the transcription system for the Icelandic Parliament, Althingi. The purpose of the system is to automatically transcribe parliamentary speeches, which are then manually edited by in-house editors and published on the Althingi website. The objective of the analysis was to gain insight on the system performance with respect to editing effort. The analysis focused on determining the relationship of the word edit distance with respect to edit time per word and the real-time factor of edits. The secondary focus was to evaluate the contribution of punctuation-related errors and the quality of automatically produced transcriptions as perceived by the editors.

The study shows that editors currently take more time to edit automatic transcripts than manual transcripts, as both observed measures, ET/W and RTF, were higher. Further analysis shows that high WED negatively affects edit time. Improving the automatic transcription to the level exhibited by manual transcription process would require lowering DB WED\_wp to 12.6\%. This conclusion is further supported by selectively looking at results for low error rate speeches, as the corresponding ET/W and RTF are similar to the resultant values for the fully manual transcription process. Despite the comments from editors, analyses do not show punctuation having a significant contribution to edit time. Further analysis is warranted before a decisive conclusion on this matter is reached. Based on the fact that only 11\% of transcriptions received a bad grade, the Althingi in-house editors were satisfied with the experimental transcription system and its integration. Further teasing out and grouping of the factors would provide useful insights into what else an ASR integration to an existing transcription process requires.

\bibliographystyle{IEEEbib}
\bibliography{refs}

\end{document}